\documentclass[iop]{emulateapj}
\usepackage[dvips]{epsfig,color}
\usepackage{amssymb,amsmath}

\newcommand{\eg}{{\it e.g.,}}
\newcommand{\ie}{{\it i.e.,}}
\newcommand{\etal}{{\it et al.}}

\newcommand{\ignore}[1]{\relax}
\newcommand{\dif}{\ensuremath{\: \mathrm{d}}}

\newcommand{\dopt}{\ensuremath{\delta_{\rm opt}}}
\newcommand{\dsim}{\ensuremath{\delta_{\rm sim}}}

\shorttitle{Extended Systems and Softening}
\shortauthors{Barnes}

\begin{document}

\title{Comparing Extended System Interactions with Motions in Softened
Potentials}
\author{Eric I. Barnes}
\affil{Department of Physics, University of Wisconsin --- La Crosse,
La Crosse, WI 54601}
\email{barnes.eric@uwlax.edu}

\begin{abstract}

Using an $N$-body evolution code that does not rely on softened
potentials, I have created a suite of unbound interacting cluster
pair simulations.  The motions of the centers of mass of the
clusters have been tracked and compared to the trajectories of point
masses interacting via one of four different softened potential
prescriptions.  I find that the relationship between the impact
parameter of the cluster interaction and the point-mass softening
length that best approximates each cluster's center-of-mass motion
depends on the adopted prescription.  In general, the range of allowed
softening lengths grows roughly linearly with the impact parameter,
but zero softening is acceptable in the majority of situations.  In an
$N$-body simulation that adopts a fixed softening length, such
relationships lead to the possibility of two-body effects, like
dynamical friction, being either larger or smaller than the
corresponding cluster situation.  Further consideration of more
specific $N$-body situations leads estimating that a very small
fraction of point-mass encounters experience two-body effects
significantly different than those of equivalent clusters.

\end{abstract}


\section{Introduction}\label{intro}

It is common for $N$-body integration schemes to rely on softened
potentials to speed up calculations.  Small impact parameter, two-body
encounters in a Newtonian potential can drive timesteps to very small
values in some $N$-body implementations \citep[\eg][{\small NBODY2}]{a01}.
Softened potentials relax this behavior by capping the maximum
acceleration any particle can experience, thereby placing a
lower limit on the timestep.  From another perspective, softened
potentials allow an $N$-body system to behave more like a
collisionless system, where two-body effects are absent.  Minimizing
strong scattering has led other $N$-body codes to adopt softened
potentials \citep[\eg][{\small GADGET-2}]{s05}.

One way of bringing some physical reality to the idea of using
softened potentials is to imagine that each particle in a simulation
is actually a cluster of unresolved particles.  This work investigates
the interactions of binary clusters of particles relying strictly on
Newtonian potentials (but note that throughout this work these
clusters are not bound to one another).  This is made possible
by using the graphics processing unit (GPU) enhanced version of
Aarseth's {\small NBODY6} code \citep{a03,na12}.  For comparison
purposes, all simulations have been completed on a multi-core,
64-bit, dual-processor machine with an NVIDIA Quadro K2000 GPU.  The
local configuration used in this work utilizes {\small GCC}
4.4.7 and {\small CUDA} 4.0.8 compilers.

I have focused on four different softened potential prescriptions to
describe the two-body interactions that model the motions of each
cluster's center-of-mass.  The first three belong to a class of
prescription sometimes referred to as `Plummer-like' softenings.
These are the type of softening implemented in {\small NBODY2}.
Prescription one is described by
\begin{eqnarray}
\boldsymbol{F}_1 & = & -\frac{G m_1 m_2}{(r+\delta_1)^2} \hat{r},
\nonumber \\
U_1 & = & -\frac{G m_1 m_2}{r+\delta_1},
\end{eqnarray}
where $r$ is the separation distance and $\hat{r}$ is the direction
from one mass to the other.  The gravitational constant $G$ is set to
unity in this work and the masses $m_1=m_2=1/2$ when referring to the
masses of clusters.  
Prescription two is described by
\begin{eqnarray}
\boldsymbol{F}_2 & = & -\frac{G m_1 m_2}{(r^2+\delta_2^2)} \hat{r},
\nonumber \\
U_2 & = & \frac{G m_1 m_2}{\delta_2}\left[ \arctan{\left(
\frac{r}{\delta_2} \right)} - \frac{\pi}{2} \right].
\end{eqnarray}
Note that this is not a useful potential, as it does not reduce to the
Newtonian point-mass potential for $\delta_2=0$.  However, the force
expression is Newtonian in that limit, allowing particle trajectories
to be integrated.  The third Plummer prescription is described by
\begin{eqnarray}
\boldsymbol{F}_3 & = & -\frac{G m_1 m_2 r}{(r^2+\delta_3^2)^{3/2}}
\hat{r}, \nonumber \\
U_3 & = & -\frac{G m_1 m_2}{\left( r^2 + \delta_3^2 \right)^{1/2}}.
\end{eqnarray}
The choice of these three prescriptions is not intended to be
exhaustive but merely representative of the kinds of softenings one
might adopt.

The fourth prescription is a version of the `spline' softening idea
discussed by \citet{hk89} in the context of smoothed particle
hydrodynamics.  Based on a smoothing kernel function described in
\citet{ml85}, {\small GADGET-2} smooths two-body interactions by
assuming that the interacting particles are actually distributed
masses ``smeared'' by the smoothing kernel.  The smoothing kernel
complicates the potential and force expressions far beyond those above
for the Plummer prescriptions; the details of their derivation may be
found in \citet{syw01}.  Generally, the spline softening approach uses
three different approximations to the force, depending on the
separation of the particles.  This behavior allows spline softened
forces to exactly equal Newtonian values when particles are adequately
separated.  With Plummer softenings, the softened force is never
exactly Newtonian, even though the difference becomes computationally
negligible with adequate separation.  In {\small GADGET} codes, the
softening length is different than the smoothing parameter, and I have
adopted the same multiplicative factor to connect them.  As
highlighted in \citet{s05}, this guarantees that the spline softened
potential caused by a point mass when $r=0$ is the same as that for a
Plummer softened potential.

Much of the basic physics of cluster-cluster interactions has been
investigated in-depth over the past several decades.  Early analytical
attempts at describing the impact of a passing mass on a spherical
system \citep[\eg][]{s58,a65} informed more numerical studies of the
phenomena \citep[\eg][and references therein]{rn79,aw85}.  It is not
the goal of this paper to revisit or duplicate the types of studies
represented by these previous works.  They are important to the
present work only in the sense that they provide guideposts for
the behavior of the clusters in the {\small NBODY6} simulations.  As
each cluster's center-of-mass in this work is intended to
represent a point mass present in collisionless $N$-body simulations,
I focus on initial conditions that lead to hyperbolic orbits.  The
clusters here are intended to slide past one another without the
possibility of becoming bound.  I envision these clusters being
what one would see if a microscope could be used to magnify the inner
structure of individual masses in an $N$-body simulation that utilizes
force softening.

Previous discussions of softening in $N$-body simulations can be
grouped into different categories.  Some researchers have focused on
optimizing softening lengths to accurately reproduce forces and/or
minimize computational requirements \citep{r98,aetal00,petal03}.
Others have discussed more fundamental issues related to the
appropriateness of Plummer-like softening formulas \citep{di93,g96} or
re-interpreting softening prescriptions as smoothing operations
\citep{b12}.  The work presented here likewise investigates the impact
that smoothed forces can have on the dynamics of an $N$-body system.
My addition to this chain is the investigation of how adequately
softened point-mass motions can model extended cluster center-of-mass
motions.  In a practical sense, the results of this work will
hopefully inform the choices of future $N$-body simulators.  For
example, if comparable dynamical accuracy can be achieved with smaller
computational effort, perhaps a simple Plummer softening prescription
would be more attractive than the (relatively) more complicated spline
prescription.

To begin, I present numerical details of the initialization procedures
(\S~\ref{ics}) and the evolutions (\S~\ref{evos}).  Discussions of the
main analyses of the dynamical evolutions and the key results
that follow form the majority of Section~\ref{anr}.  Finally, I
present a summary of the techniques and outcomes in
Section~\ref{conc}.

\section{Numerical Details}\label{num}

\subsection{Initial Conditions}\label{ics}

Clusters are composed of $N_c$ particles, where $N_c=2^{\eta}$ and $10
\le \eta \le 14$, but $\eta=13$ ($N_c=8192$) has been adopted as the
standard value.  The positions of the particles in one cluster are
randomly chosen according to a spherical density distribution
with maximum radius $R_c=1$.  This work investigates clusters with
equilibrium Plummer distributions \citep{p11} with scalelength equal
to one-tenth of $R_c$ and Gaussian distributions.  The scalelength of
the Gaussian distributions used is generally set to $R_c/\sqrt{5}$;
this makes the Gaussian clusters less centrally concentrated than the
Plummer clusters.  Particles are given equal masses such that the
total mass of a single cluster is $M_c=1/2$.  For Plummer clusters,
thermal velocities are isotropic and speeds are chosen from the
Maxwellian speed distribution.  An upper limit to particle speed is
given by $3v_{r,{\rm rms}}$ at the appropriate radial location.  With
this choice, a single cluster is initially closer to virial
equilibrium than if the upper limit were set to the escape speed
\citep[$2v_{r,{\rm rms}}$, \eg][]{ahw74}.  Particle velocities in
Gaussian clusters are chosen to be isotropic and so that the cluster
is in virial equilibrium.  Dynamical evolutions of isolated, single
clusters show a very modest mass loss; $\Delta M_c/M_c \approx 1$ per
cent for $N_c=1024$ and $\Delta M_c/M_c \approx 0.5$ per cent for
$N_c=16384$.  For single cluster {\small NBODY6} evolutions, particles
escape if they reach distances of 10 half-mass radii from the center
of the cluster.  The isolated cluster evolutions show evidence of some
outer envelope expansion.  The radius that contains 95 per cent of the
mass of the cluster increases by approximately 5 per cent during an
evolution lasting 10 single-cluster crossing times.  For comparison,
the radius containing 75 per cent of the mass increases by less than 1
per cent.

Once a single cluster has been initialized, its center-of-mass
location is moved from the origin and a mirror image of the cluster is
created (different particle distributions can be chosen for the two
clusters; see Section~\ref{repeat}).  Particle velocities are simply
mirrored as well.  Unless otherwise noted, initial center-of-mass
separations in the $x$-direction $\Delta x_{\rm init}$ are equal to
four times the initial cluster radius $R_c$, while there is no offset
in the $y$-direction.  The majority of simulations I have performed
have center-of-mass separations in the $z$-direction (impact
parameters) $\Delta z_{\rm init}$ that range between one and five
times $R_c$, but values up to $40 R_c$ have been investigated.

With all particle initial locations specified, the total potential
energy $U_{\rm total}$ is determined.  The self-potential energy of a
cluster is the potential energy determined between all particles in
that cluster.  The total potential energy is the sum of the two
cluster self-potential energies and an inter-cluster potential energy
determined between particles in the different clusters.  The speeds of
each cluster's center-of-mass are defined as fractions of $U_{\rm
total}$,
\begin{equation}\label{relvel}
v_{\rm CM, initial} = \sqrt{Q U_{\rm total}},
\end{equation}
where $0.2\le Q \le 1.5$ for the majority of the simulations discussed
here.  To provide some context to the $Q$ values, I have compared the
average thermal speed of particles in a Plummer cluster to the initial
relative center-of-mass speed for clusters.  The average thermal speed
is that of an equilibrium Plummer sphere and is invariant to different
cluster initial conditions.  Varying the initial center-of-mass
separation changes the inter-cluster potential energy, but by a very
small amount relative to the self-potential energies.  As a result,
the total potential energy is roughly equal to twice the
self-potential of an equilibrium Plummer sphere, and the ratio of
thermal to relative center-of-mass speeds is essentially just a
function of $Q$,
\begin{equation}
\frac{v_{\rm CM, relative}}{v_{\rm thermal}} \approx 2.2 \sqrt{Q}.
\end{equation}
Specifically, $Q$ values of 0.2, 0.5, 0.8, and 1.5 result in
ratios of 1.0, 1.6, 2.0, and 2.7, respectively.  The less-concentrated
nature of the Gaussian clusters makes their potential energies smaller
than those of Plummer clusters, with a corresponding reduction in the
center-of-mass speeds of Gaussian versus Plummer clusters at a given
$Q$. 

The initial center-of-mass velocities are chosen to lie in the
$x$-direction.  Finally, particles in each cluster have the
appropriate center-of-mass velocity added to their pre-existing,
thermal velocity.  With all velocities assigned, an evolution
timescale is calculated from,
\begin{equation}
t_{\rm dyn}=\sqrt{\frac{(2M_c)^5}{(2|E_{\rm total}|)^3}},
\end{equation}
where $E_{\rm total}$ is the sum of $U_{\rm total}$, the
center-of-mass kinetic energies of the clusters, and the thermal
kinetic energies of the clusters.

\subsection{Simulations}\label{evos}

The {\small NBODY6} code advances all particles without softening
their interactions.  All simulations have been evolved for a minimum
of ten timescales.  For Gaussian and Plummer clusters with $Q=0.2$
(lower $v_{\rm CM, relative}$ values), simulations have been extended to
twenty and fifteen timescales, respectively.  Visualizations of
cluster interactions show that these values provide sufficient time
for clusters to reach a point of minimum center-of-mass separation and
then clearly separate.  With the very different initial conditions
involved in the binary cluster evolutions, the escaping particle
criterion has been changed from the {\small NBODY6} default behavior.
For interacting cluster simulations, particles escape only if they
reach distances 50 times the quantity $(2M_c)^2/U_{\rm total}$.  This
allows simulations with large center-of-mass separations (impact
parameters up to 40 $R_c$) to be performed without artificially
removing particles.  With this new escape criterion, very few
particles are lost from systems.

During each evolution, particle positions and velocities are compiled
every $0.1 t_{\rm dyn}$.  The total number of
particles in the system is reported at every output along with a list
of particle identification numbers.  By comparing lists of identifying
particle numbers with the initial list, the cluster membership of each
particle can be tracked.  From this data, I calculate the
centers-of-mass positions and velocities and then the thermal
velocities of cluster particles.  With those basics, I then calculate
the thermal and center-of-mass kinetic energies of each cluster, the
self-potential energies of each cluster, and the inter-cluster
potential energy as functions of time.  Total and individual cluster
angular momenta and average cluster radius values are also tabulated
as functions of time.

The results of these simulations match those reported in earlier works
dealing with similar systems and initial conditions
\citep[\eg][]{drra87}.  Interactions with impact parameters $\Delta
z_{\rm init} > 2R_c$ are elastic in the sense that there is no
change in either cluster's center-of-mass kinetic energy during
symmetric time intervals centered on the point of closest approach.
For smaller impact parameters that lead to overlap of the systems at
closest approach, the center-of-mass kinetic energies decrease.  The
self-potential energies of clusters generically increase.  This agrees
with the increase of average radius of each cluster during an
interaction.  Small impact parameter cases increase the average
radii by roughly 15 per cent around the point of closest approach,
while larger impact parameters produce more modest changes, around 5
per cent.  Since the clusters remain self-gravitating, their expansion
is coupled with a decrease in thermal kinetic energy \citep{s58}.  For
larger impact parameter situations, the increase in self-potential has
the same magnitude as the decrease in thermal kinetic energy.  For
smaller impact parameters, the loss in center-of-mass kinetic energy
further boosts the self-potential change.

Cluster angular momentum changes are generally very small.  No angular
momentum component changes by more than 1 per cent of the initial
magnitude of the system angular momentum during the time interval
surrounding the point of closest approach.  Another perspective on the
size of these changes is that similarly sized changes occur when
individual particles are lost from a cluster.  Absent
finite-particle-number variations, the clusters initially have zero
angular momentum about their centers-of-mass.  In the simulations with
the smallest impact parameters, tidal interactions are able to torque
them slightly.  The small angular momenta acquired in these cases are
oriented in the same direction as the initial total angular momentum.

\section{Analysis \& Results}\label{anr}

\subsection{Determining Softening Behaviors}

For any given set of initial conditions, the motions of cluster
centers-of-mass have been matched to the motions of point masses
interacting via the softened potentials discussed in
Section~\ref{intro}.  Treating a softening length as a free parameter,
the non-linear minimization routine amoeba \citep{press94} is used to
determine what value of $\delta$ produces the best representation of
the center-of-mass motion.  The figure of merit used by amoeba is a
combination of least-squares deviations.  The softened motion starts
from the simulated cluster initial conditions, and an adaptive
timestep Runge-Kutta scheme (implemented in the {\small IDL} routine
{\tt lsode}) integrates softened motion between the times at which
simulated data (center-of-mass positions and velocities) exists.
Given the $M$ time values at which ``data'' values of $x(t_i)$,
$y(t_i)$, and $z(t_i)$ produced by {\small NBODY6} exist, {\tt lsode}
advances the point mass locations and velocities from $t_i$ to
$t_{i+1}$.  The results shown and discussed here use a relative
tolerance of $10^{-7}$; fourth- and fifth-order Runge-Kutta steps
result in relative differences smaller than the tolerance. Tests using
a tolerance down to $10^{-11}$ have also been performed, but do not
lead to any appreciable differences.  Updating the time index $i$,
the former final conditions become initial conditions and are
recorded.  In this way, a set of $M$ corresponding softened ``model''
position and velocity values are produced.  From these $M$ pairs of
values, figures of merit that quantify differences between data and
model values can be formed.  The main figure of merit used in this
work is formed by comparing position values.  For each cluster, the
$M$ differences between center-of-mass and softened point $x$, $y$,
and $z$ positions are calculated.  In each direction, and for each
cluster, the average of the squared difference values is used to form
an rms deviation.  The two cluster rms values are then averaged in
each direction.  Finally, the figure of merit is formed by summing the
directional rms values.  In this way, the figure of merit reflects the
average positional deviation of the model from the data over the
entire simulation.  The optimal softening length \dopt\ is that which
produces the best agreement (smallest figure of merit) between the
motion of the two clusters' centers-of-mass and the softened point
particles.

\begin{figure}
\includegraphics{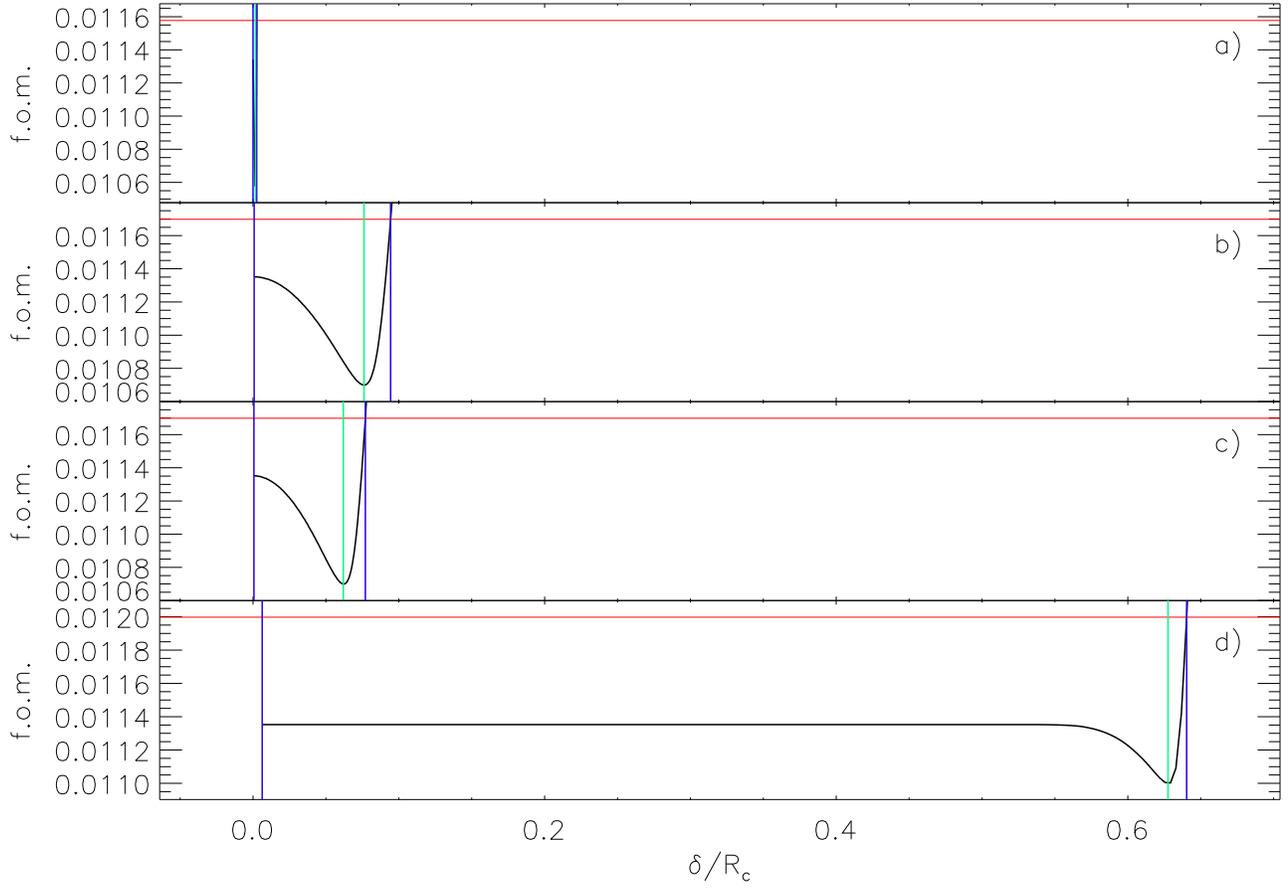}
\caption{Panels a, b, c, and d show figure-of-merit curves for
softening prescriptions one, two, three, and four, respectively.
These curves are based on a Plummer cluster simulation with
$N_c=8192$, $Q=0.2$, $\Delta x_{\rm init}=4R_c$, $\Delta z_{\rm
init}=2R_c$.  The thin horizontal line indicates the limiting
figure-of-merit value, minimum plus $10^{-3}$.  The thick, solid
vertical lines mark the low and high range values for the softening
value, while the thick, dashed vertical line indicates the \dopt\
value determined by the amoeba routine.  Due to the rather large range
imposed by the prescription 4 \dopt\ value and the rather small \dopt\
value for prescription 1, it is nearly impossible to see the
prescription 1 curve.  However, viewing that curve individually
reveals it to behave very similarly to the prescription 2 and 3
curves.  Note that for prescription 4 (panel d) the minimum is
actually nearer the high end of the range.  This is a generic behavior
for prescription 4 (see, for example, Figure~\ref{doptvballp}).
\label{suncert}}
\end{figure}

\begin{figure}
\includegraphics{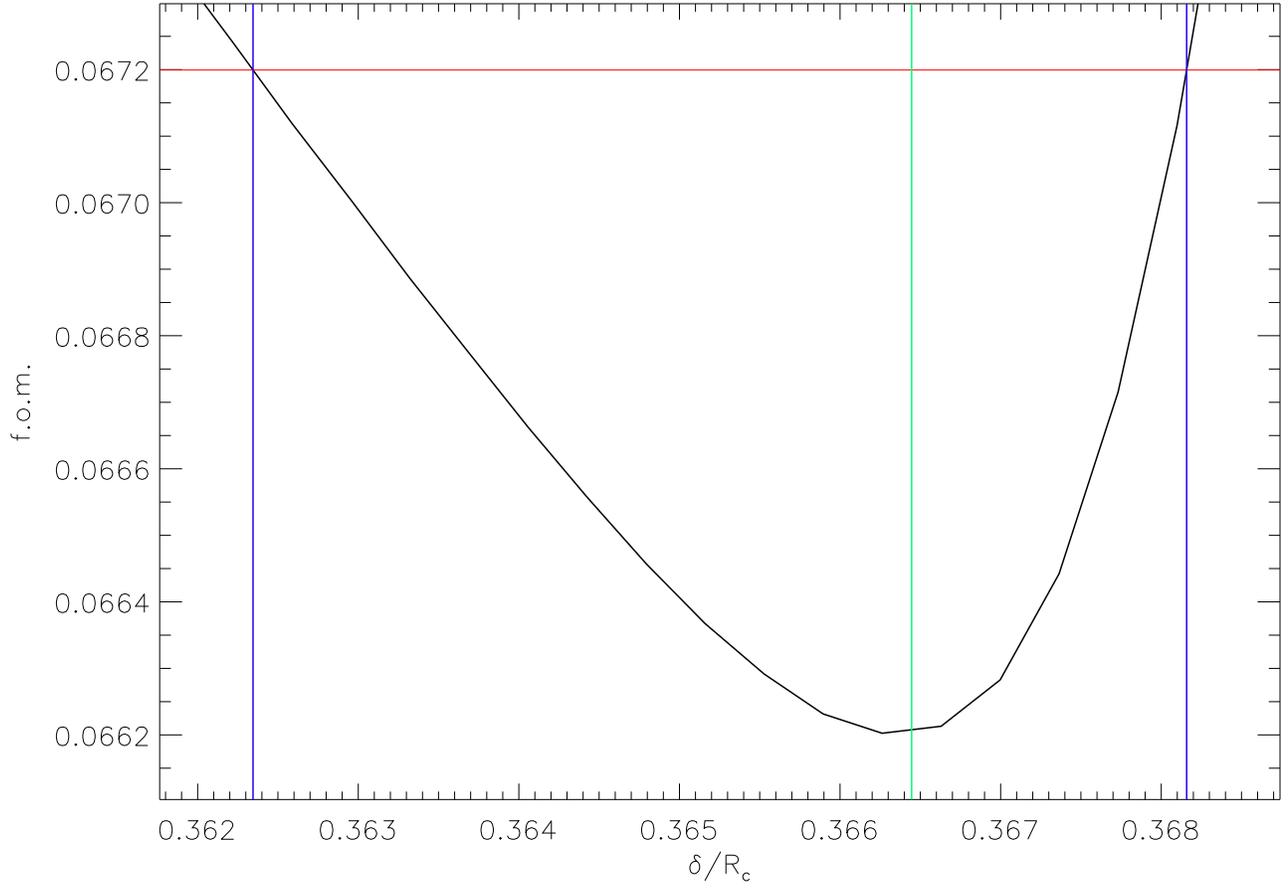}
\caption{A single, prescription four figure-of-merit curve for a
Plummer cluster interaction with $N_c=8192$, $Q=0.5$, $\Delta x_{\rm
init}=4R_c$, and $\Delta z_{\rm init}=R_c$.  Contrast the
quadratic-like behavior here with the one-sided trough seen in
Figure~\ref{suncert}d.  Interactions with the smallest impact
parameters investigated here tend to enforce lower limits to the
softening length regardless of other initial conditions or cluster
density profile.
\label{suncert2}}
\end{figure}

With a \dopt\ value, the figure-of-merit landscape is searched near
the minimum to provide uncertainty values.  I have adopted a value of
$10^{-3}$ as the relevant figure-of-merit change, as that is roughly
the size of the error in the location of a cluster center-of-mass.  In
sum, a given \dopt\ produces the best fit for a given model, and
$\Delta \delta$ is determined by finding the two neighboring softening
values that produce figures of merit that are $10^{-3}$ greater than
the minimum value.  Figure~\ref{suncert} shows a representative
example of the information used to determine uncertainties.  In
general, prescriptions one, two, and three form quadratic-like
figure-of-merit curves.  Prescription four tends to produce the kind of
one-sided valleys seen in Figure~\ref{suncert}.  This is unsurprising
as the spline softening approach deals with ranges of softening
values.  What is mildly surprising is the fact that the prescription
four \dopt\ values always tend to be near the high-end of the
uncertainty range.  However, it is important to keep in mind that the
difference in figure-of-merit values is extremely small between
$\delta \approx 0$ and \dopt.  In all but the smallest impact
parameter situations, one can safely assume a softening length of zero
with prescription four.  For contrast with Figure~\ref{suncert},
Figure~\ref{suncert2} focuses on the behavior of the prescription four
figure-of-merit curve for a simulation where Plummer clusters have
$\Delta z_{\rm init}=R_c$.  Very small softening values are clearly
ruled out in these kinds of interactions.

This choice of figure of merit is straightforward but not unique.  I
have varied the calculation of \dopt\ to use differences between
velocities rather than positions.  Resulting values of \dopt\ fall
within the uncertainties derived from the position-based calculations.
All remaining discussions of \dopt\ values and/or figure-of-merit
values refer to the position-based approach.

\subsection{Comparison of Softening Prescriptions}

Two questions motivated this investigation.  First, do particles
interacting through softened potentials really behave like interacting
clusters?  Second, is any one prescription better than the others?  I
tackle the second question first in this section.  However, note that
this exercise is not determining an optimal form of softening
potential, merely comparing several extant potentials.

For any simulation, four \dopt\ values are determined, one for each
prescription.  I use the corresponding best-fit figure-of-merit values
as a comparison tool.  For 25 Plummer cluster simulations ($N_c=8192$,
$\Delta x_{\rm init}=4R_c$, $R_c \le \Delta z_{\rm init} \le 5R_c$, $0.2
\le Q \le 5.0$), prescription 1 provides the best match 15 times,
prescription 2 is best 4 times, prescription 3 is best 1 time, and
prescription 4 is best 5 times.  However, the differences between the
values for the different prescriptions are generally very small.  For
example, in its 15 ``wins'', the figure of merit for prescription 1
is, on average, smaller than its competitors by about $2\times
10^{-4}$.  For the other prescriptions, their margins of victory are
roughly 100 times smaller.  Since these differences are smaller than
the error in determining the center-of-mass position, I conclude that
all of the prescriptions provide comparable levels of match.

For a similar set of Gaussian cluster simulations, prescriptions 1 and
4 are evenly split; prescription 1 is best 8 times, prescription 4 is
best 10 times.  The softness of the Gaussian clusters make it
more difficult to determine how much better each prescription is.  In
simulations with $Q \le 0.5$ and $\Delta z_{\rm init}=R_c$, the
clusters experience strong tidal distortions.  In the $Q=0.2$ case,
the two clusters actually merge during their encounter, making its
analysis here worthless.  Even though that case has been ignored here,
it gives one the idea that figure-of-merit values will be generally
larger than in the Plummer simulations, with correspondingly larger
differences seen.  In the 8 ``wins'' for prescription 1 (several of
which are small impact parameter cases), the average figure of merit
is about 0.01 smaller than that of its competitors.  If the
smallest impact parameter cases are removed, that value drops to about
$10^{-3}$.  Prescription 4 tends to win in higher impact parameter
situations, so its margin of victory of $10^{-5}$ is rather firm.
While there is more hand-waving in this analysis, I again conclude
that the prescriptions are basically equivalent in their ability to
explain the center-of-mass motions of these clusters.

\subsection{Optimal Softening Lengths}

\begin{figure}
\includegraphics{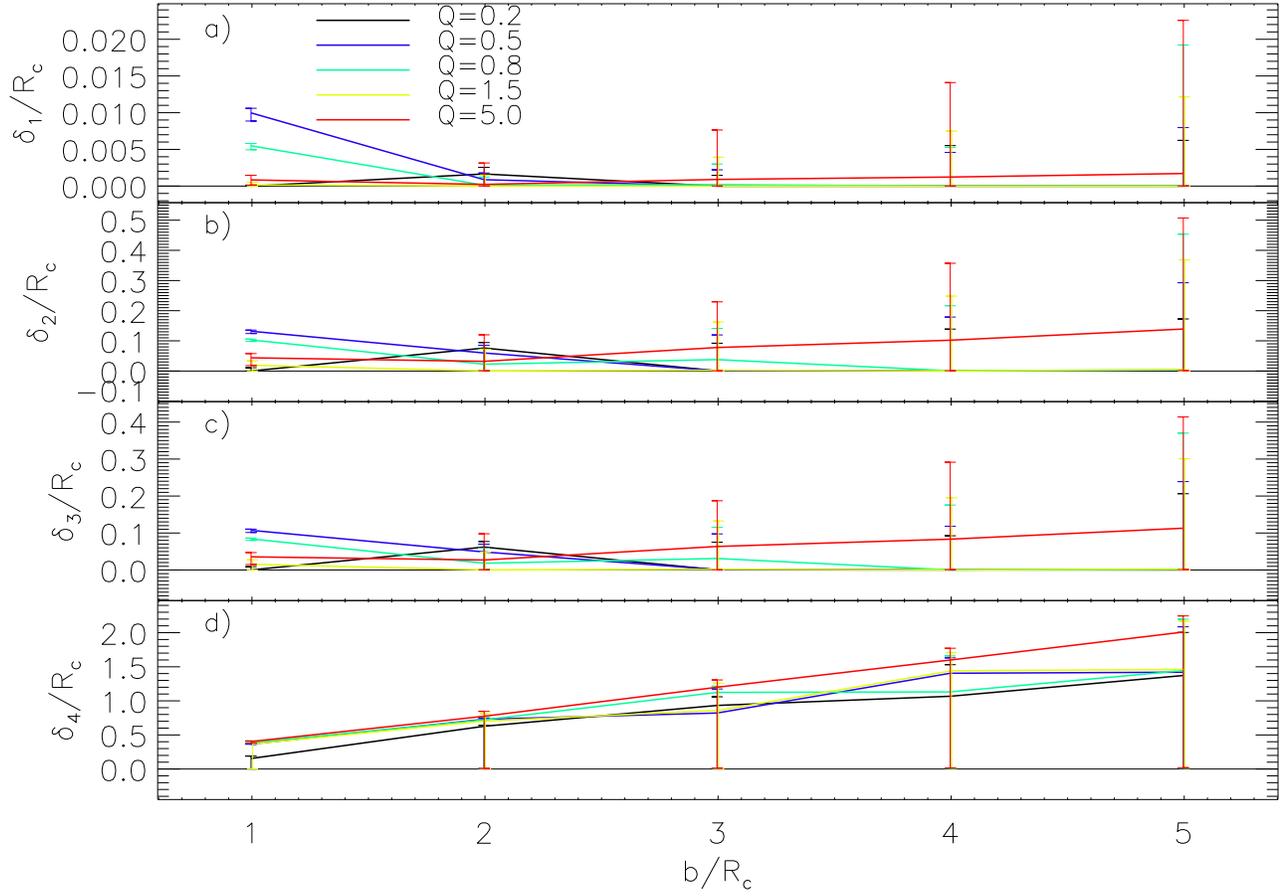}
\caption{The relationships between the optimal softening length
$\dopt$ and impact parameter for the four softening prescriptions in
simulations of Plummer clusters with $N_c=8192$.  Lines connect the
\dopt\ values, while the errorbars indicate the ranges for those
values.  Each line represents results from simulations with different
$Q$ values.  Panels a, b, c, and d correspond to prescriptions one,
two, three, and four, respectively.  Softening lengths and impact
parameters are scaled by the initial cluster radius $R_c$.
Uncertainty ranges grow with impact parameter.  For all but the
smallest impact parameter cases, softening lengths of zero are
reasonable.
\label{doptvballp}}
\end{figure}

\begin{figure}
\includegraphics{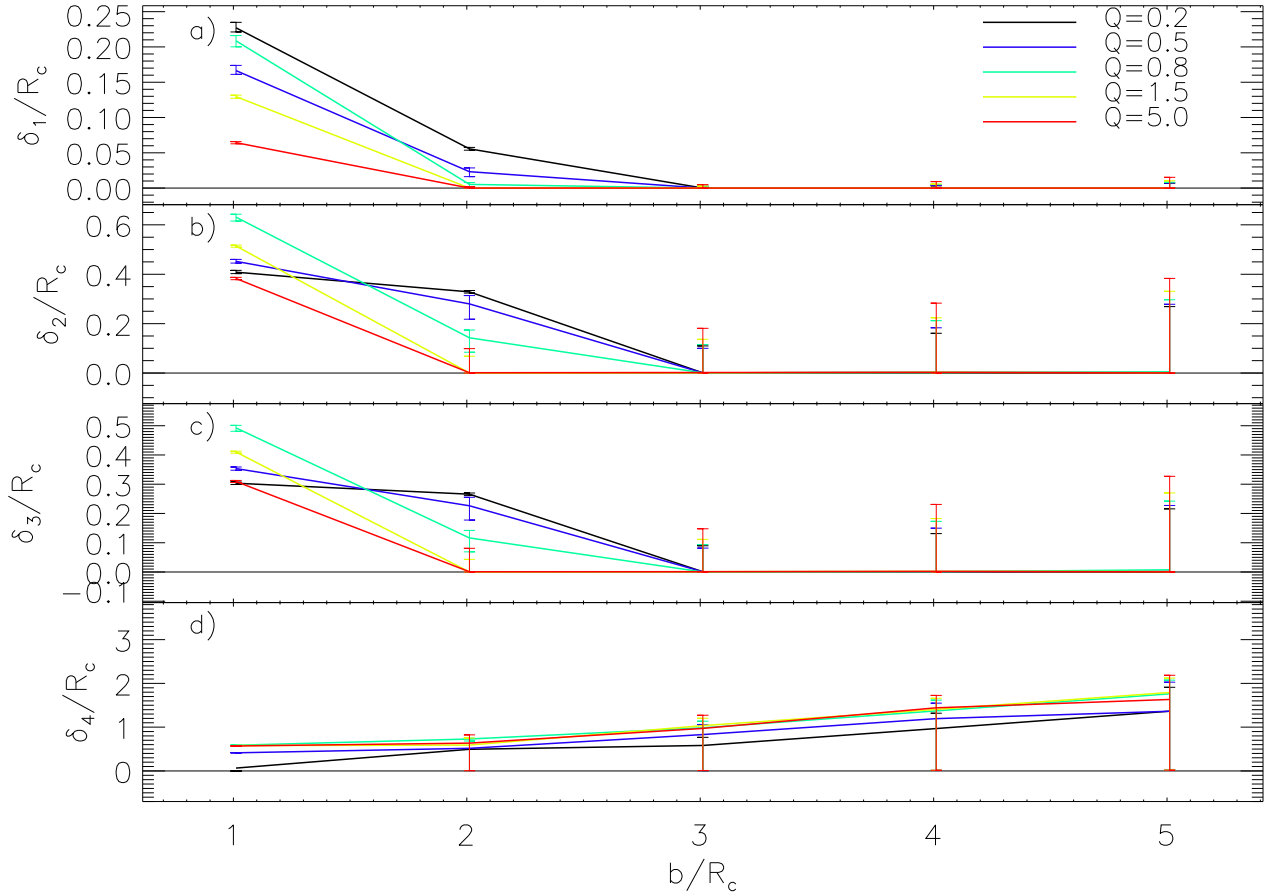}
\caption{The relationships between the optimal softening length
$\dopt$ and impact parameter for the four softening prescriptions in
simulations of Gaussian clusters with $N_c=8192$.  Panels and
linestyles are the same as those in Figure~\ref{doptvballp}.  The
$Q=0.2$, $b/R_c=1$ results are not reliable as those simulations lead
to cluster convergence.  There is a trend for smaller impact
parameter cases to result in larger \dopt\ values.  As with the
Plummer clusters, softening lengths of zero are reasonable for larger
impact parameter cases.
\label{doptvballg}}
\end{figure}

Figure~\ref{doptvballp} shows the results of determining the optimal
softening length \dopt\ using the four softening prescriptions given
initially-equal-radius Plummer clusters with differing impact
parameters in simulations where $Q$ takes on the values 0.2, 0.5, 0.8,
1.5, and 5.0.  A similar plot for Gaussian clusters is presented in
Figure~\ref{doptvballg}.  Overall, the trend is for \dopt\ values to
decrease as the impact parameter increases.  As discussed above, it is
also typical to see non-zero lower limits to $\delta$ for smaller
impact parameters.

\begin{figure}
\includegraphics{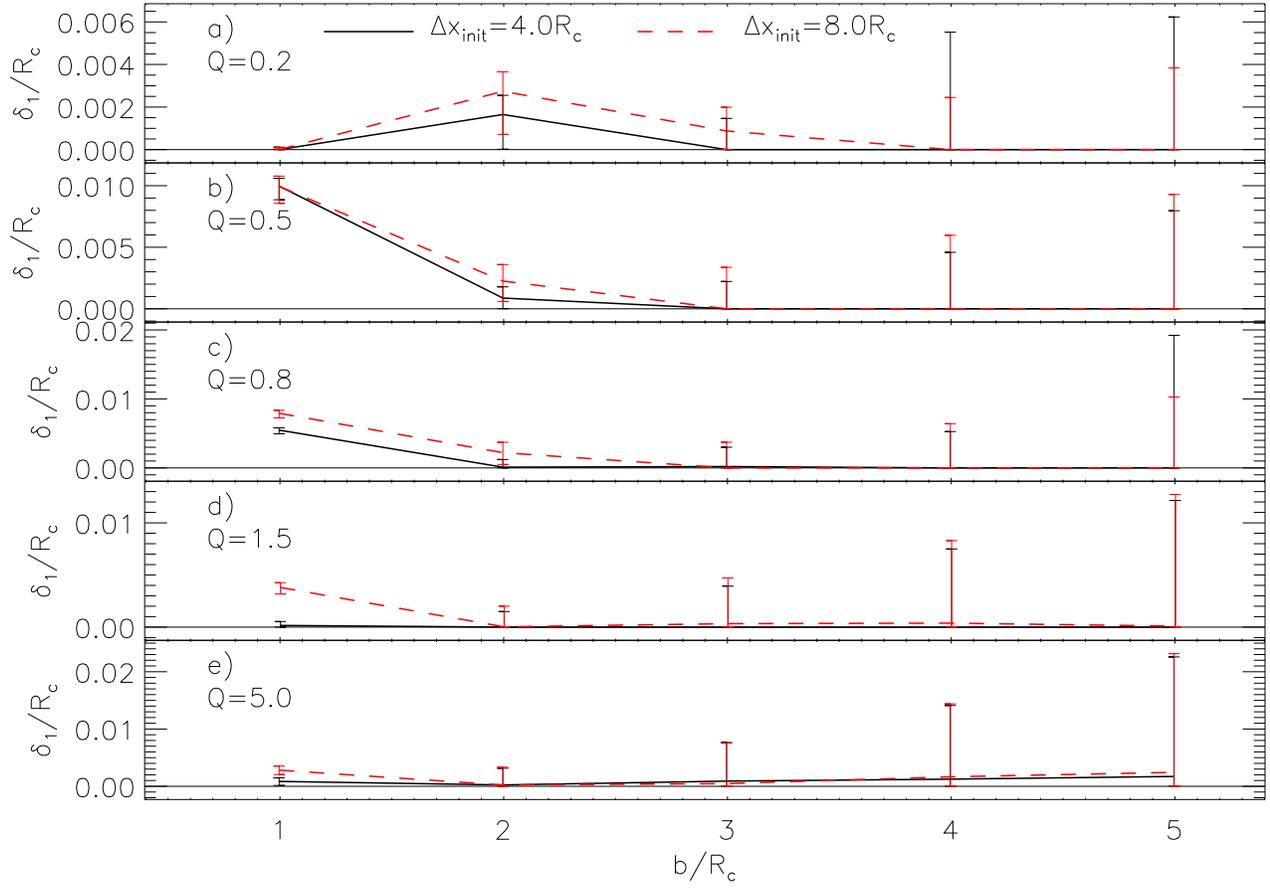}
\caption{Comparisons between prescription one softening lengths given
different initial separations in the $x$-direction.  Panels a, b, c,
d, and e correspond to results from simulations with $Q=0.2$, $Q=0.5$,
$Q=0.8$, $Q=1.5$, and $Q=5.0$, respectively.  All simulations have
$N_c=8192$.  Versions of this plot using information from different
prescriptions and/or cluster density profile show similar levels of
agreement.
\label{xoffcomp}}
\end{figure}

In an attempt to filter out the impact of some other parameter values,
I have created additional simulations with different initial
separations and different cluster particle numbers.  Simulations with
initial separations in the $x$-direction twice the standard value
produce $\dopt(b)$ curves comparable to those in
Figure~\ref{doptvballp}.  Figure~\ref{xoffcomp} shows how the
$\dopt(b)$ curves differ between Plummer cluster $\Delta x_{\rm
init}=8$ and $\Delta x_{\rm init}=4$ simulations with $N_c=8192$.  For
clarity, only prescription one \dopt\ values are shown.  The level of
agreement shown here is typical of what is seen for the other
prescriptions and for results from Gaussian clusters.

\begin{figure}
\includegraphics{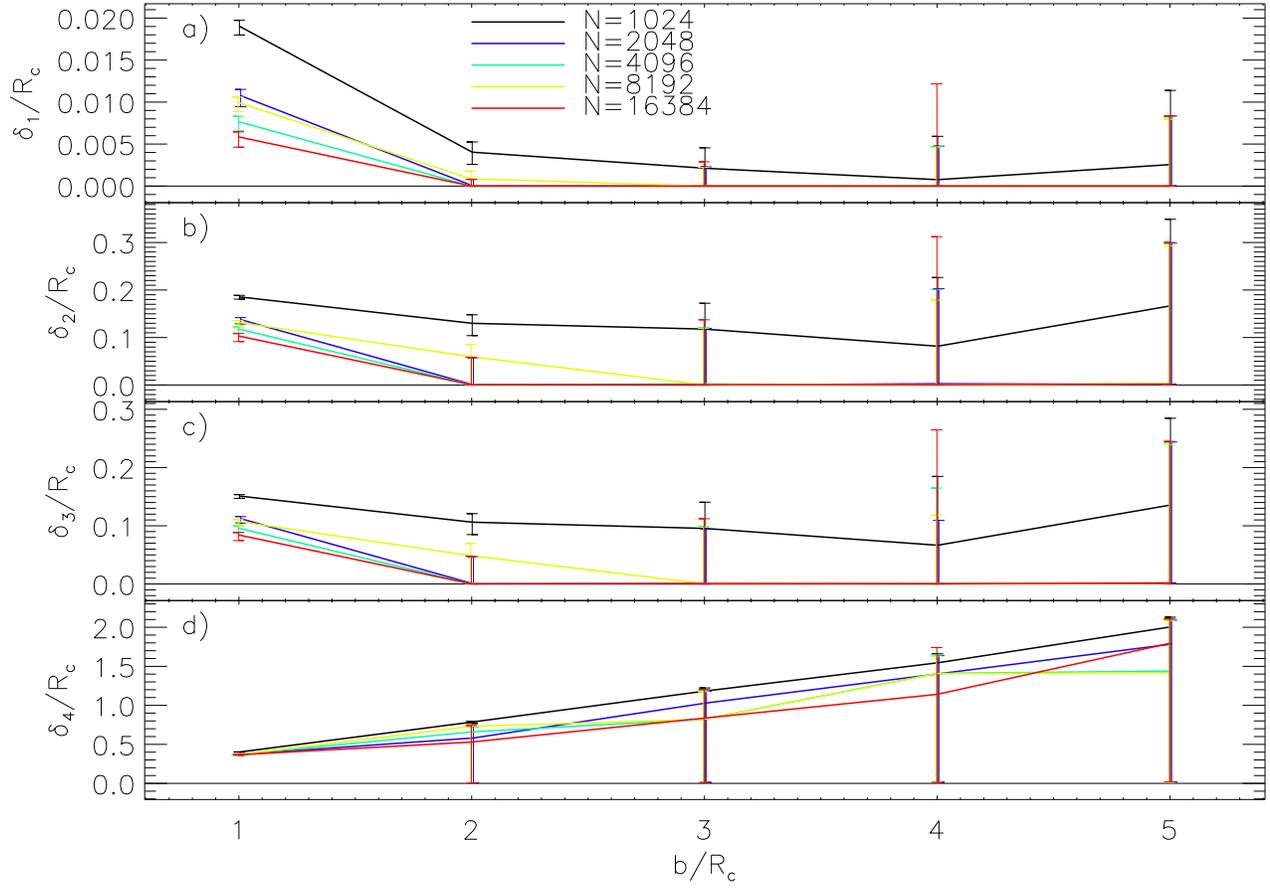}
\caption{Comparisons between softening values given different cluster
particle numbers.  Panels a, b, c, and d correspond to the different
softening prescriptions.  The various linestyles correspond to
different particle numbers $1024 \le N_c \le 16384$.  In general,
\dopt\ values and uncertainties are consistent as long as $N_c \ge
2048$.
\label{numcomp}}
\end{figure}

The results of varying particle numbers using Plummer clusters with
$Q=0.5$ are shown in Figure~\ref{numcomp}.  The panels isolate the
different softening prescriptions, and the different lines in each
correspond to the various $N_c$ values.  Overall, the clusters with
the lowest resolution ($N_c=1024$) result in the largest values of
\dopt.  For $N_c \ge 2048$, the values of \dopt, uncertainties in
\dopt, and trends in \dopt\ versus $b$ are essentially the same.

\subsection{Results of Non-optimal Softening}\label{results}

To investigate the impact of these $\dopt(b)$ relationships, imagine
an $N$-body simulation with a specified softening length \dsim.
Depending on the adopted softening prescription, one would like to
choose \dsim\ so that the majority of interactions would accurately
describe the center-of-mass motion of two clusters of particles.  

For encounters with $\dsim \ne \dopt(b)$, the point masses will follow
trajectories that do not match those of each cluster's
center-of-mass with the same impact parameter.  To clarify the
outcome of such encounters, further imagine simulating three pairs of
point masses with the same impact parameter $b_0$.  In simulation A,
$\delta_{\rm sim A} > \dopt(b_0)$; in simulation B, $\delta_{\rm sim
B} < \dopt(b_0)$, while in simulation C, $\delta_{\rm sim C} =
\dopt(b_0)$.  I imagine the C pair as centers-of-mass for two
clusters, since they follow the same trajectories.  With a larger
softening length, the pair in simulation A will experience a smaller
change in velocity than those in simulation C.  In general, 
point-mass interactions with $\dsim > \dopt(b)$ result in motions
that underestimate the change in velocity relative to true cluster
interactions.  Conversely, the pair in simulation B will experience a
larger change in velocity compared to those in simulation C;
interactions with $\dsim < \dopt(b)$ generically result in 
point-mass motions that overestimate the change in velocity relative
to cluster interactions.  If softened point-mass motions are intended
to represent cluster center-of-mass motions, a single softening
parameter is insufficient to correctly handle different impact
parameters.

It is standard to break the velocity change experienced by interacting
point masses into components parallel and perpendicular to the initial
direction of relative motion.  The parallel component is related to
the dynamical friction a particle will experience while the
perpendicular component relates to the relaxation time for the motion.
In an $N$-body simulation with a single fixed softening length \dsim,
encounters where $\dsim >(<) \dopt(b_0)$ lead to weaker (stronger)
dynamical friction and relaxation effects.  Based on the ranges of
$\dopt$ values presented above, very few situations will lead to
an overestimation of dynamical friction/relaxation; $\delta \approx 0$
is within the uncertainties in nearly all simulations.

The question becomes, what is the likelihood for a particle in a
softened $N$-body simulation to have an encounter with any given
impact parameter?  I take the $N$-body system to be a time-independent
background volume density and investigate what a test particle moving
through that system would encounter.  The $N$-body system is spherical
with radial extent $\mathcal{R}$.  For a test particle on a radial
orbit, I estimate the likelihood of an encounter by calculating the
projected density of particles in the system.  This projected density
is a function of a distance $R$ measured in a plane perpendicular to
the velocity of the particle on a radial orbit.  The $R$ value is just
the impact parameter for the test particle and a ring of system
particles centered on the origin.  The projected density integrated
around an annulus provides the number of system particles within a
range of impact parameters.  For a uniform density system, this
annular density has a maximum value for $R_{\rm
max}/\mathcal{R}=1/\sqrt{2}$.  However, more realistic
centrally-concentrated systems have maximum annular densities that
occur much closer to the center.  As a simple example, consider a
cored power-law volume density with asymptotic behaviors similar to
the Plummer law,
\begin{equation}\label{powerrho}
\rho=\frac{\rho_0}{(1+r/r_0)^5},
\end{equation}
where $\rho_0$ is the maximum number density of system particles and
$r_0$ is a density profile scalelength.  In this situation, $R_{\rm
max}/\mathcal{R} \approx 0.05$ when $\mathcal{R}=10 r_0$.  As a
result, roughly 80 per cent of particles in such a system have
$b/\mathcal{R} \la 0.25$ and about 15 per cent have $b/\mathcal{R} \la
0.05$.  A Navarro-Frenk-White \citep[][NFW]{nfw97} volume density
produces a similar value of $R_{\rm max}/\mathcal{R}$ but has a more
slowly decaying annulus density curve.  Roughly 80 per cent of
particles in such a system have $b/\mathcal{R} \le 0.55$ with about 10
per cent having $b/\mathcal{R} \la 0.05$.  Steepening the outer volume
density profile to a Hernquist model \citep{h90} shrinks $R_{\rm
max}/\mathcal{R}$ to approximately 0.03 and 80 per cent of particles
have $b/\mathcal{R} \la 0.35$.  Roughly 10 per cent of particles
have $b/\mathcal{R} \la 0.03$ in this case.

For test particles on circular orbits, the approach is different.  At
any point along the orbit, the distribution of impact parameters
encountered by the particle is given by the projected background
density function in the vicinity of the orbit.  For a set of orbit
radii $0.01 \le R_0 \le \mathcal{R}$, I have integrated the projected
density function over a circular region with radius $b$ centered on
the orbit.  For the most centrally-located circular orbits
investigated, the fraction of system particles with $b \le R_0$ grows
to 10 per cent when $R_0/\mathcal{R} \approx 0.04$ for the
Plummer-like cored power-law volume density.  With a Hernquist system
profile the values are very similar, but for an NFW volume density
this level is not reached until $R_0/\mathcal{R} \approx 0.08$.  For
larger orbits with values of $0.1 \le R_0/\mathcal{R} \le 0.3$,
roughly 25 per cent of system particles have $b \le R_0$, when the
background is the Plummer-like density.  This percentage increases to
about 35 per cent for an NFW volume density.  These sample
calculations suggest that in an isotropic $N$-body system, with a mix
of point masses on radial and circular orbits, it is not unreasonable
to expect roughly 10 per cent of impact parameters to be small
compared to the radius of the system.

I now draw connections between the cluster radius $R_c$, the size of a
system in an $N$-body simulation $\mathcal{R}$, and the softening
length.  Taking the average density of one point mass spread over a
cluster to be the same as the background system density $\rho$ at a
radius $r$, the cluster radius can be written as
\begin{equation}
R_c^3 = \frac{3}{4\pi \rho(r)}.
\end{equation}
The minimum cluster radius is set by the maximum density.  Adopting
the density of Equation~\ref{powerrho},
\begin{equation}
R_c=\left( \frac{3I}{N_{\rm tot}}\right) ^{1/3} r_0,
\end{equation}
where $N_{\rm tot}$ is the total number of particles in the simulation
and 
\begin{displaymath}
I = \int_0^{\mathcal{R}/r_0} \frac{u^2}{(1+u)^5} \dif u.
\end{displaymath}
For $5 \le \mathcal{R}/r_0 \le 50$, $I \approx 0.08$.  This leads to 
\begin{equation}\label{rcr}
R_c \approx 0.1 N_{\rm tot}^{-1/3} \mathcal{R}.
\end{equation}
For an NFW background density, the right-hand side of
Equation~\ref{rcr} is approximately twice as big.  As I am being
rather rough in this discussion, the factor of two is ignored and no
distinction is made between the background density profiles.

For $10^5 \le N_{\rm tot} \le 10^7$, commonly realized values,
Equation~\ref{rcr} leads to $R_c \approx 10^{-3} \mathcal{R}$.
Following the argument in \citet{petal03}, the $N$-body simulation
should not have softened accelerations larger than the mean-field
acceleration at the system edge.  This leads to the requirement that
\begin{equation}\label{pdelt}
\dsim \ge \frac{\mathcal{R}}{\sqrt{N_{\rm tot}}}.
\end{equation}
Combining Equations~\ref{rcr} and \ref{pdelt},
\begin{equation}\label{rcdelt}
R_c \la 0.1 \dsim N_{\rm tot}^{1/6}.
\end{equation}
For the same range of $N_{\rm tot}$ mentioned above,
Equation~\ref{rcdelt} leads to $\dsim/R_c \ga 1$.

Combined with the previous estimates of fractions of particles
encountered by test particles, approximately 10 per cent of
interactions will have impact parameters less than about 40 $R_c$.  As
this value is much larger than the range of impact parameters that
have been discussed so far ($b/R_c \le 5$), I have also run a limited
number of binary cluster simulations with larger impact parameters
($b/R_c \le 40$).  The \dopt\ values for these simulations do not show
monotonic trends, but the uncertainty ranges do.  The lower
limits of the ranges remain near zero, but the upper limits
grow roughly linearly with $b$.

For prescription one, the upper limits to $\delta$ stay below
$R_c$ until $b \approx 20 R_c$.  For prescriptions two and three,
$\delta/R_c$ upper limits reach values of one at $b \approx 10
R_c$.  With prescription four, the upper limit to $\delta/R_c
\approx 1$ when $b \approx 2 R_c$.  The inference I draw from these
results is that adoption of the Power argument for choosing a
simulation softening length will lead to underestimating violent
relaxation and relaxation effects in a very small percentage of
interactions, those with very small impact parameters.

One must be careful to not push this very simple analysis too
far.  In an actual situation with clusters interacting as parts of a
much larger system further complications arise.  In particular, tidal
forces can affect the rates of separation of each cluster's
center-of-mass \citep[\eg][Figure 1]{petal16}.  A quick calculation
suggests that the results presented here would not change dramatically
in the presence of a background potential.  Giving the background
system a constant density (as in the core of an pseudo-isothermal
halo) leads to a differential gravitational acceleration field that is
constant,
\begin{equation}
\left.\frac{\dif g}{\dif r}\right|_{\rm background} = \frac{G
\mathcal{M}}{\mathcal{R}^3},
\end{equation}
where $\mathcal{M}$ is the mass of the background system and $G=1$
will be assumed from here on.  I note that a cuspy distribution with a
background density that is inversely proportional to radius produces
no differential accelerations.  Taking the background system to be
composed of $N$ other clusters, $\mathcal{M}=N M_c$.  With the
relationships developed earlier in this section, I estimate
\begin{equation}
\left.\frac{\dif g}{\dif r}\right|_{\rm background} = \frac{M_c}{(10
R_c)^3},
\end{equation}
where $N=10^6$ has been adopted, in agreement with the other
relationships.  For comparison, I take the differential gravitational
acceleration due to a neighboring cluster to be that of a point mass
(since the clusters should not overlap),
\begin{equation}
\left.\frac{\dif g}{\dif r}\right|_{\rm cluster} =
\frac{-2M_c}{(\alpha R_c)^3},
\end{equation}
where $\alpha R_c$ is the distance from the center of the cluster.
Taking a cluster to have these two contributions to tidal forces, we
see that the relative strengths are comparable for $\alpha \approx
10$.  As $\alpha < 10$ at the points of closest approach for the
simulations presented here, I estimate that background tidal effects
should be a small perturbation.  With an eye toward future work, note
that \citet{petal16} highlight that the details of the orientations of
cluster orbits with respect to the larger system (\ie\ radially or
tangentially biased) are important for accurate determination of tidal
effects.  Such an expansive investigation is far beyond the scope of
the present work.

\subsection{Repeated Interactions}\label{repeat}

The expansion of clusters that accompanies interactions leads to the
question of how cluster size affects the softening behavior.
Specifically, if earlier interactions had expanded a cluster, will a
subsequent interaction with an unexpanded cluster still be modeled by
the same $\dopt(b)$ relationship?  I note that adopting this as a
physical picture requires one to think of clusters as self-gravitating
systems that remain too small to be resolved in a larger-scale
simulation.

I have taken two approaches towards dealing with this situation.
These simulations all start with the same center-of-mass initial
conditions as the previously discussed same-radius simulations.  In
the first approach, one of the clusters is allowed to have an outer
radius twice as big as the other.  The Plummer profiles of the two
clusters are defined by the same scalelength, but one can be populated
out to a larger radius.  These different-radius-same-scalelength
simulations model situations where previous encounters have tidally
extended the envelope of a cluster without affecting its core.  The
second approach assumes that the cluster core has also been
strongly affected, with one cluster having twice the Plummer
scalelength and outer radius of the other.

\begin{figure}
\includegraphics{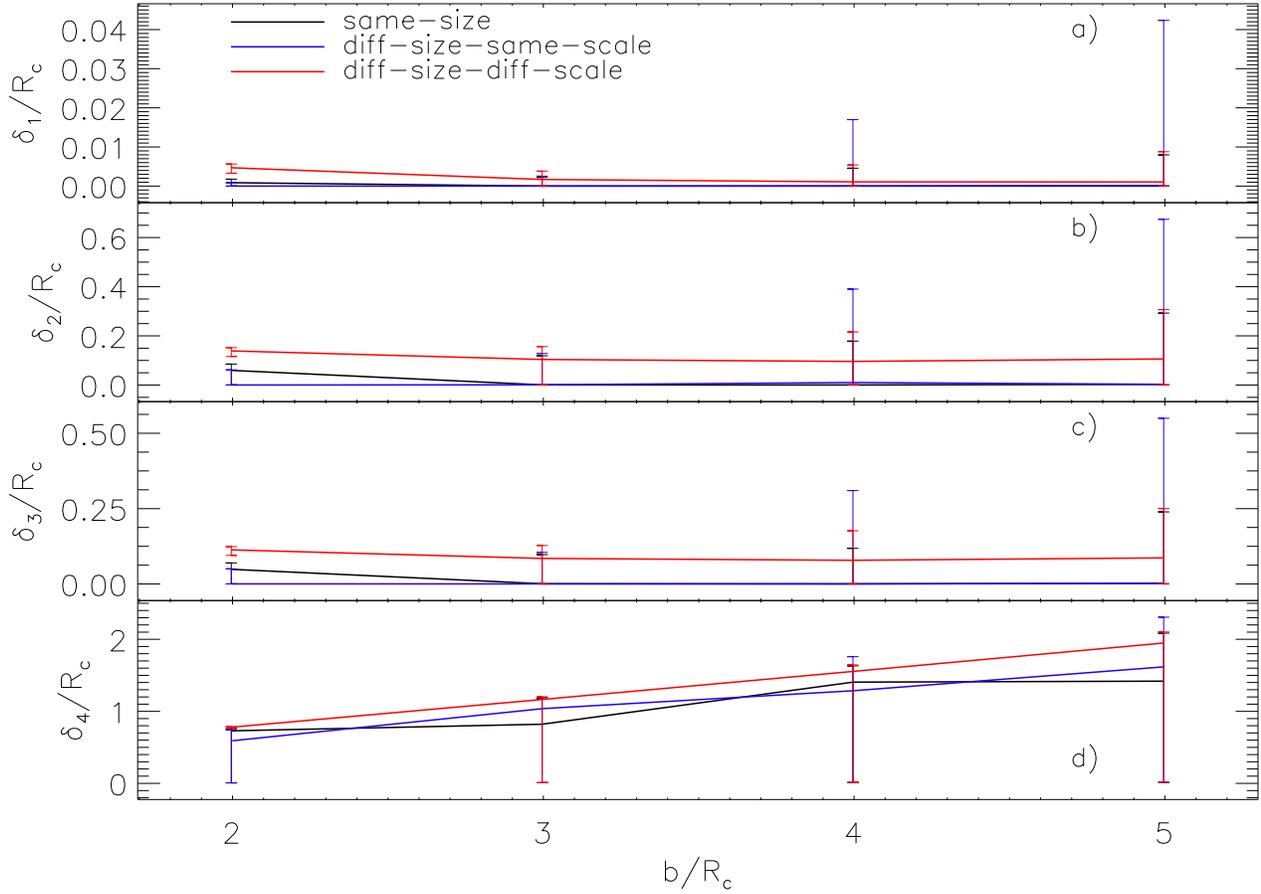}
\caption{Comparisons between $\dopt(b)$ relationships for $N_c=8192$,
$Q=0.5$, $\Delta x_{\rm init}=4R_c$ simulations in which the Plummer
clusters have initially (i) same outer radii (`same-size' line), (ii)
different outer radii but same density scalelengths
(`diff-size-same-scale' line), and (iii) different outer radii and
different Plummer density scalelengths (`diff-size-diff-scale' line).
No different-size simulations have been created with $b=1$, as they
lead to excessive cluster overlap.  In simulations with differences,
the length in one cluster is twice that in the other cluster.  For
example, in the `diff-size-same-scale' simulations, one cluster has
twice the outer radius of the other.  Panels a, b, c, and d correspond
to prescriptions one, two, three, and four, respectively.  In general,
interactions with a ``puffed up'' cluster (larger outer radius and
larger scalelength) produce larger \dopt\ values.  However,
differences in \dopt\ values are smaller than the uncertainties in all
but the smallest impact parameter cases.
\label{sizecomp}}
\end{figure}

Differences between the three families of simulations are evident in
their energy and radius-change histories.  Energies in the
different-radius-same-scalelength simulations behave very much like
those in the previously discussed same-radius cases (see
Section~\ref{evos}).  In the different-radius-and-scalelength
simulations, energy values are now very different between the
clusters, but the self-potential energy changes are still roughly
of the same magnitude as the thermal kinetic energy changes in each
cluster.  Clusters in all simulations expand as a result of their
interactions.  As with the same-radius situations, each cluster
experiences a comparable fractional increase in radius in the
different-radius-same-scalelength cases.  For the
different-radius-different-scalelength situations, the larger cluster
experiences a smaller fractional increase in size compared to the
small cluster but maintains the largest size throughout the
simulation.  Despite these differences, the $\dopt(b)$ relationships
derived from the different-radius simulations are basically identical
to those from the same-radius simulations.  Figure~\ref{sizecomp}
shows how the $\dopt(b)$ relationships for same-radius,
different-radius-same-scalelength, and
different-radius-different-scalelength simulations with $N_c=8192$
compare. 

If the point masses of an $N$-body simulation were replaced with
clusters that have initially identical radii, encounters would quickly
result in a distribution of cluster sizes and scalelengths.  From the
simple cases investigated here, it is reasonable to assume that
the different cluster sizes will not appreciably alter the conclusions
based on the same-size encounters.

\section{Summary}\label{conc}

I have modeled the motions of interacting binary cluster
centers-of-mass with a set of softened potential prescriptions.  These
kinds of simulations have a substantial history in the literature,
where they have been used to investigate galaxy-galaxy interactions in
galaxy clusters \citep[\eg][]{rn79, aw85}.  The simulations discussed
here are unique in that they cover a very different region of
interaction parameter space and no softening is employed in the
evolution of the clusters.

The main inferences drawn from these simulations are:
\begin{itemize}

\item Cluster centers-of-mass follow trajectories that can be
well-described by point masses moving under the influence of softened
potentials.

\item The softening prescriptions studied here provide matches to
cluster center-of-mass motions that are comparable.  Differences in
the qualities of the fits provided by each prescription are generally
smaller than the uncertainty in a cluster center-of-mass position.

\item Optimal softening lengths can be found for any cluster-cluster
interaction by minimizing discrepancies between cluster
centers-of-mass trajectories and softened point-mass motions.  For the
Plummer prescriptions, optimal softening length values are largest for
the closest encounters.  For the spline prescription, the optimal
softening length values grow nearly linearly with impact parameter.
Independent of prescription, uncertainties in the optimal values grow
with the impact parameter.  Softening lengths near zero fall within
the uncertainties for all but the closest encounters.

\item Given a prescription, the relationships between softening length
and impact parameter for cluster interactions are robust against
significant changes to initial conditions: cluster particle number,
initial separation distance, relative velocity, and internal
structure.

\item Softened point-mass interactions correctly describe cluster
centers-of-mass motions when the softening length is optimal.  For
softenings less than the optimal value, the point masses
experience more relaxation and dynamical friction effects than
interacting clusters would.  With softenings larger than optimal,
point masses experience weaker two-body interactions than clusters
would.

\item A negligible fraction of the interactions experienced by
point masses in an $N$-body simulation will have impact parameters that
place them in the regime of experiencing weaker relaxation and
dynamical friction effects than would clusters.  However, collapse
simulations that involve varying numbers of particles with given
impact parameters will most likely be affected differently, possibly
more severely, than the static situations discussed here.

\end{itemize}

\section*{Acknowledgements}

Many thanks to UW-L student Jacob Gloe for his efforts towards
optimizing the graphics-processing-unit-enabled {\small NBODY6} code
for local use and suggesting the idea for the investigation.  I am
indebted to an anonymous referee for numerous scientific and editorial
suggestions that have strengthened this article.  Thanks also to an
anonymous colleague for pushing me to find an error in an earlier
version of the analysis software.  This work was partially supported
by a University of Wisconsin --- La Crosse Faculty Research award.

\end{document}